\documentclass[onecolumn]{aastex61}
\usepackage{graphicx}
\usepackage{soul}

\revised{\today}

\shorttitle{Improved Meridional Flow Measurements}
\shortauthors{Mahajan et al.}

\begin{document}

\title{Improved Measurements of the Sun's Meridional Flow and Torsional Oscillation from Correlation tracking on MDI \& HMI magnetograms}

\correspondingauthor{Sushant S. Mahajan}
\email{mahajan@astro.gsu.edu}

\author{Sushant S. Mahajan}
\affil{Department of Physics \& Astronomy, Georgia State University, Atlanta, GA, USA.}

\author{David H. Hathaway}
\affil{W.W. Hansen Experimental Physics Laboratory, Stanford University, Stanford, CA, USA.}

\author{Andr\'es Mu\~noz-Jaramillo}
\affil{Southwest Research Institute, Boulder, CO, USA.}
\affil{High Altitude Observatory, Boulder, CO, USA}
\affil{National Solar Observatory, Boulder, CO, USA}

\author{Petrus C. Martens}
\affil{Department of Physics \& Astronomy, Georgia State University, Atlanta, GA, USA.}


\begin{abstract}
The Sun's axisymmetric flows, differential rotation and meridional flow, govern the dynamics of the solar magnetic cycle and variety of methods are used to measure these flows, each with its own strengths and weaknesses. Flow measurements based on cross-correlating images of the surface magnetic field have been made since the 1970s which require advanced numerical techniques that are capable of detecting movements of less than the pixel size in images of the Sun. We have identified several systematic errors in addition to the center-to-limb effect that influence previous measurements of these flows and propose numerical techniques that can minimize these errors by utilizing measurements of displacements at several time-lags. Our analysis of line-of-sight magnetograms from the {\em Michelson Doppler Imager} (MDI) on the ESA/NASA {\em Solar and Heliospheric Observatory} (SOHO) and {\em Helioseismic and Magnetic Imager} (HMI) on the NASA {\em Solar Dynamics Observatory} (SDO) shows long-term variations in the meridional flow and differential rotation over two sunspot cycles from 1996 to 2020. These improved measurements can serve as vital inputs for solar dynamo and surface flux transport simulations.

\end{abstract}

\keywords{ Correlation Tracking, Meridional Flow, Torsional Oscillation, Center-to-limb effect}


\section{Introduction} \label{sec:intro}

The Sun's axisymmetric flows, differential rotation (DR) and meridional flow (MF) play key roles in virtually all models of the Sun's magnetic dynamo \citep{2010LRSP....7....3C}. The differential rotation (variation in rotation rate with both latitude and depth) stretches the radial and latitudinal components of the magnetic field in the longitudinal direction - thereby increasing the field strength and changing the field direction. The meridional flow transports the radial and longitudinal components of the magnetic field in the latitudinal direction - thereby building up the polar fields on the surface \citep{1989SoPh..124....1S,2010ApJ...717..597J} and annihilating oppositely directed longitudinal fields across the equator in the solar interior. Thus, measurements of the meridional flow are important for the prediction of the build up of  polar fields which in turn are good predictors for the next solar cycle \citep{2013ApJ...767L..25M}. 

Measurements of these axisymmetic flows have been made with a wide variety of methods. \cite{1859MNRAS..19...81C} measured the positions of sunspots to determine the latitudinal variation in rotation rate through the limited range of sunspot latitudes. Since those earliest measurements using the only known tracer at the time (sunspots), many more measurements have been made using a variety of tracers and/or methods. Tracers include: sunspots, magnetic features (network magnetic elements), Doppler features (supergranules), and intensity features (granules). Measurement methods (beside feature tracking) include direct Doppler (spectrographic) and helioseismic measurements. \cite{2000SoPh..191...47B} has compared and contrasted these various methods. Each has its own set of advantages and disadvantages. This is particularly true for measurements of the meridional flow, primarily due to the intrinsic low amplitude of this global flow.

Sunspot tracking is limited in both time and space. Sunspots only appear in the lower latitudes and at times don't appear at all. Most other methods have much better coverage in space (latitude) and time but present other problems.

Direct Doppler measurements of the meridional flow \citep{1996Sci...272.1306H, 2010ApJ...725..658U} require an accurate determination of the convective blue shift signal and its dependence on local magnetic fields. This signal is stronger than the meridional flow signal and is similar in spatial characteristics - making the meridional flow measurement via its direct Doppler signal very difficult.

Feature tracking the Doppler features (supergranules) has an advantage in having a strong signal near the limb - and hence high latitudes, but suffers from systematic changes in the features due to line-of-sight effects. \cite{2006ApJ...644..598H} showed that these projection effects make the supergranule pattern appear to rotate faster than it actually does. The impact of line-of-sight effects on meridional flow measurements has, to our knowledge, not been investigated.

Helioseismology also has difficulty near the limb, but more importantly, exhibits a systematic center-to-limb shift in the apparent acoustic field. This effect manifests itself as a pseudo flow away from the disk center which increases as a function of center to limb distance and interferes with measurements of the meridional flow. \cite{2012ApJ...749L...5Z} accounted for it by measuring the rotation rate as a function of longitude from the central meridian at the equator and subtracting it from meridional flow measurements assuming that the effect is symmetric on the solar disk.

Feature tracking the photometric intensity features (granules) requires high spatial and temporal resolution which have only been available recently. These features are difficult to track near the limb and their short lifetimes make accurate measurements of slow flows, such as the meridional flow, quite difficult. This kind of tracking shows a ``shrinking" Sun effect (an apparent flow toward disk center) with an amplitude of $\sim 1000\ m\ s^{-1}$ at a heliocentric angle of $60^{\circ}$. \cite{2016A&A...590A.130L} investigated this effect and showed that in large parts it originates from the apparent asymmetry of granulation due to radiative transfer effects when observing at a viewing angle. However, their model of the ``shrinking Sun" effect was not accurate enough to measure the much weaker meridional flow. They also pointed out that the shrinking Sun effect was not exactly symmetric across the central meridian implying an east-west dependence.

Feature tracking the weak (network) magnetic features has the advantages of good coverage over the visible disk (but with the same difficulty near the limb as with granules and acoustic waves) coupled with long lifetimes that are advantageous for meridional flow measurements. \cite{1993SoPh..147..207K} used local correlation tracking (LCT) on magnetograms separated by a day in time. By averaging their daily measurements over two year intervals they acquired enough accuracy to show a systematic variation in the meridional flow with solar cycle phase. \cite{2010Sci...327.1350H} used a variant of the \cite{1993SoPh..147..207K} method on magnetograms obtained with SOHO/MDI \citep{1995SoPh..162..129S} at eight-hour intervals. The sheer number of observations allowed for accurate measurements over individual 27-day solar rotations. \cite{2010ApJ...722..774D} suggested that these measurements were compromised by a systematic effect due to supergranule diffusion. They argued that this diffusion would produce an apparent outflow from magnetic flux concentrations in active regions and tested their argument using a purely diffusive model with a diffuse magnetic field. \cite{2011ApJ...729...80H} investigated this possibility with a more realistic model - magnetic elements transported by evolving supergranule flows - and found that this systematic effect was too small to be measured. They further noted that the nature of the meridional flow variations with solar cycle phase produced {\em inflows} toward active latitudes, not outflows. 

Until now, a systematic center-to-limb effect was not reported in tracking network magnetic elements making it the most promising method for measuring the meridional flow. Here we describe our measurements of the meridional flow and torsional oscillation using a feature tracking method (LCT) on weak magnetic features seen in solar magnetograms. This method involves mapping data from full-disk magnetograms onto heliographic longitude-latitude maps and then cross-correlation of blocks of data between two maps separated in time by a few hours. The displacement at which two blocks separated in time show the best correlation then provides a measure of the displacement vector. Our original intent was to incorporate the daily magnetograms from NSO/Kitt Peak to extend the meridional flow measurements to cover four solar cycles but we discovered early on that several improvements to the existing LCT methods were needed (as described in Section 3). 

We, in this paper describe the discovery of a center-to-limb effect similar to the systematic flow away from disk center seen with helioseismic methods or the opposite of the ``shrinking Sun" effect seen in ganule tracking which compromises feature tracking measurements from magnetograms reported by \cite{2017ApJ...836...10L} and \cite{2018ApJ...864L...5I} and to some extent \citet{2010Sci...327.1350H}. 


Testing our LCT algorithm with synthetic magnetograms also revealed the need to improve the interpolation algorithm used for projecting magnetograms onto a heliographic grid (uniformly spaced in both latitude and longitude). Histograms of the meridional flow velocity measurements indicated: 1) the need to broaden the search for the correlation maximum to larger distances at high latitudes so as not to truncate the normal distribution (and thereby influence the value of the average), 2) the need to obtain multiple measurements in longitude with larger search windows at higher latitudes so as to cover similar areas on the surface of the Sun, and 3) the need to more accurately determine the position of the correlation maximum with fractional pixel accuracy.

In the following sections we describe the data, our displacement measurement method, discovery of systematic errors and the improvements we made. We end with a synopsis of our improved measurements of the meridional flow and torsional oscillation profile as a function of time through solar cycles 23 and 24.

\section{Data}

\subsection{MDI/HMI Magnetograms}
We obtained full disk (1024$\times$1024 pixel) line-of-sight magnetograms from the Michelson Doppler Imager (MDI) \citep{1995SoPh..162..129S} with a cadence of 96 minutes from May 4, 1996 to Jan 11, 2011 and full disk (4096$\times$4096 pixel) line-of-sight magnetograms from the Helioseismic and Magnetic Imager (HMI) \citep{2012SoPh..275..207S} with a cadence of one hour from May 20, 2010 to June 27, 2020. The MDI (HMI) magnetograms are obtained at a 60-second (45-second) cadence and averaged (by the instrument teams) over 5 minutes (12 minutes) to reduce the 5-minute oscillation signal.

These magnetograms are created from polarization measurements along the profiles of magnetically sensitive photospheric absorption lines (the \ion{Ni}{1} $676.8$ nm line for MDI and the \ion{Fe}{1} $617.3$ nm line for HMI). We use the line-of-sight magnetic measurements from both instruments and project them onto grids uniformly spaced in longitude and latitude. The methods used in projecting the data onto these grids were tested for accuracy as described in section \ref{sec:interpolation_improvement}.

\begin{figure}
\begin{flushleft}
  \centering
  \includegraphics[width=9 cm]{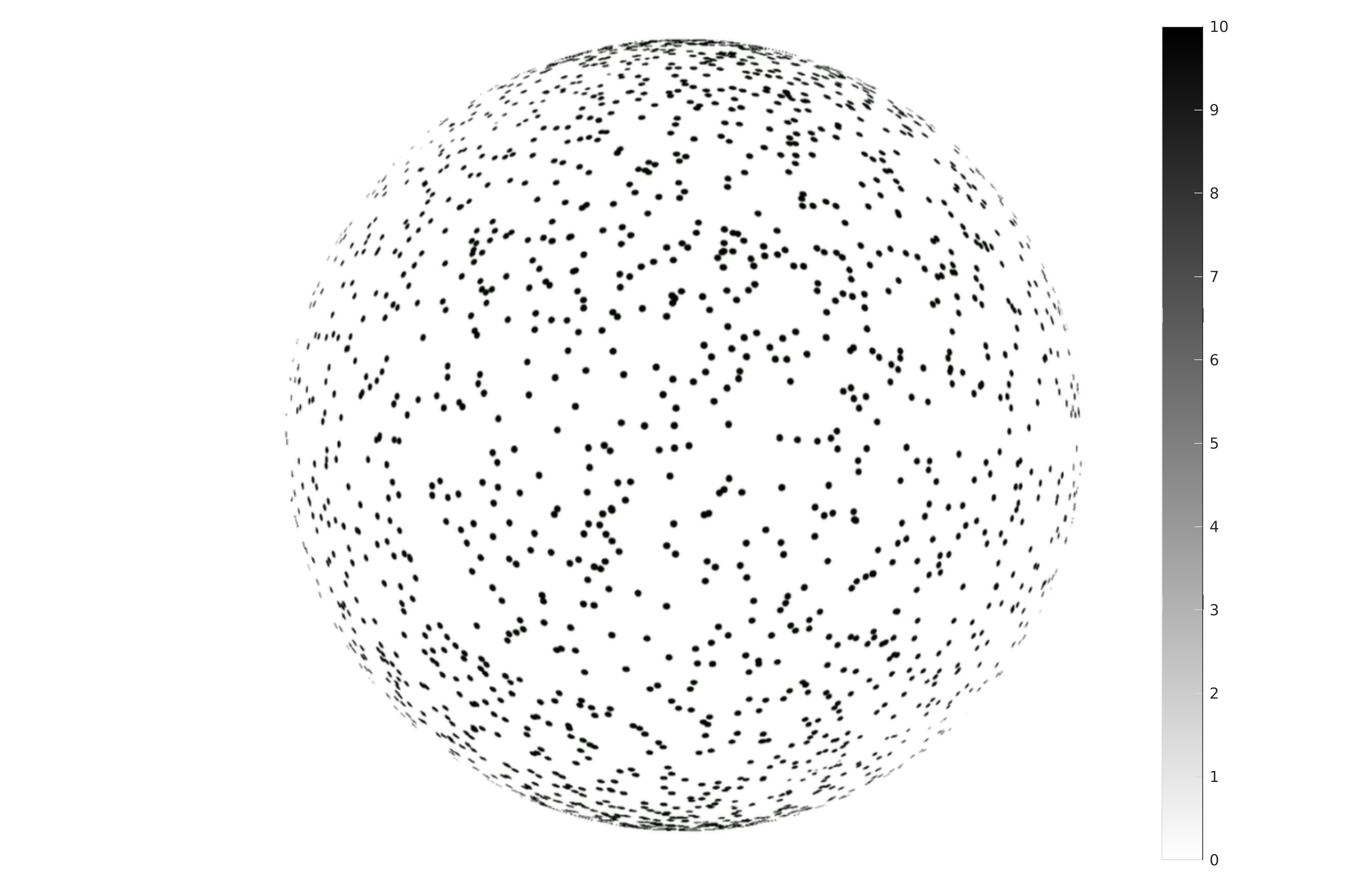}\\
  \caption{A synthetic solar magnetogram with 4000 positive polarity magnetic features produced by the Analytical Magnetogram Creator (AMC) and sampled by the CCD emulator. The magnetic field shown here is in $Gauss$.}\label{sample_image}
  \end{flushleft}
\end{figure}


\subsection{Synthetic Magnetograms for Testing}\label{sec:synthetic_data}

In order to calibrate the performance of any tracking algorithm, a ``ground-truth" dataset with known movements of features is needed. For this purpose, we developed an algorithm which allows us to create synthetic magnetograms of the Sun by placing magnetic features at predetermined locations with analytical precision.

Given the central latitude ($\lambda_{1}$) and longitude ($\phi_{1}$) of a magnetic feature, its 2D Gaussian magnetic field distribution is modelled as in Eqs.\ \ref{gaussian2d} \& \ref{great_circle_distance} below: 

\begin{equation}\label{gaussian2d}
B(\lambda,\phi)=A e^{-\frac{d^{2}}{w^{2}}}\cos(\rho+\eta)
\end{equation}

\begin{equation}\label{great_circle_distance}
d=2R_{\odot}\sin^{-1}\sqrt{\sin^{2}\bigg(\frac{\lambda-\lambda_{1}}{2}\bigg)+\cos(\lambda_{1})\cos(\lambda)\sin^{2}\bigg(\frac{\phi-\phi_{1}}{2}\bigg)}
\end{equation}

\noindent where A is the amplitude i.e. the maximum magnetic \sout{field} flux density within the feature, $\phi$ is the longitude, $\rho$ is the heliocentric angle from disk center, $\eta$ is the angle between the line-of-sight and the line joining the observer to the center of the Sun, $w$ is the width of the Gaussian and $d$ is the great circle distance between the location of the peak of the feature ($\lambda_{1},\phi_{1}$) and an evaluation point ($\lambda,\phi$).

We call this algorithm the ``Analytical Magnetogram Creator (AMC)" and its output is an analytical function which describes the radial component of magnetic flux density on the surface of the synthetic Sun. Note here that being analytical functions, magnetograms created by AMC have infinite resolution as they can be sampled at any desired resolution.

To create MDI/HMI-like magnetograms from the output of AMC, we created a Charge-Coupled Device (CCD) emulator which emulates a camera placed at a prescribed location with respect to the Sun, with a prescribed plate scale (resolution), P angle (position angle of the solar rotation axis relative to the ``vertical"), B angle (tilt of the solar rotation axis toward or away from the observer) and coordinates of the center of the solar disk. It numerically integrates the magnetic flux density within the boundaries of every image pixel by dividing each pixel into 10$\times$10 sub-pixels and evaluating the line-of-sight magnetic flux density in the total area of the pixel which is equivalent to the measured magnetic field. 


Features placed on synthetic magnetograms of the Sun can be moved precisely by a prescribed distance with insignificant numerical errors (verified empirically) because the magnetic field values come from an exact analytical formula. Such synthetic magnetograms of the Sun enable the accurate calibration of our correlation tracking algorithm. One such synthetic magnetogram generated by placing 4000 magnetic features randomly on the solar disk is shown in Fig.\ \ref{sample_image}.

\section{Improvements to the Correlation Tracking Algorithm}\label{sec:improvements}

\subsection{Latitude-Longitude interpolation used for Heliographic projection}\label{sec:interpolation_improvement}

\begin{figure}
    \centering
    \includegraphics[width = 12 cm]{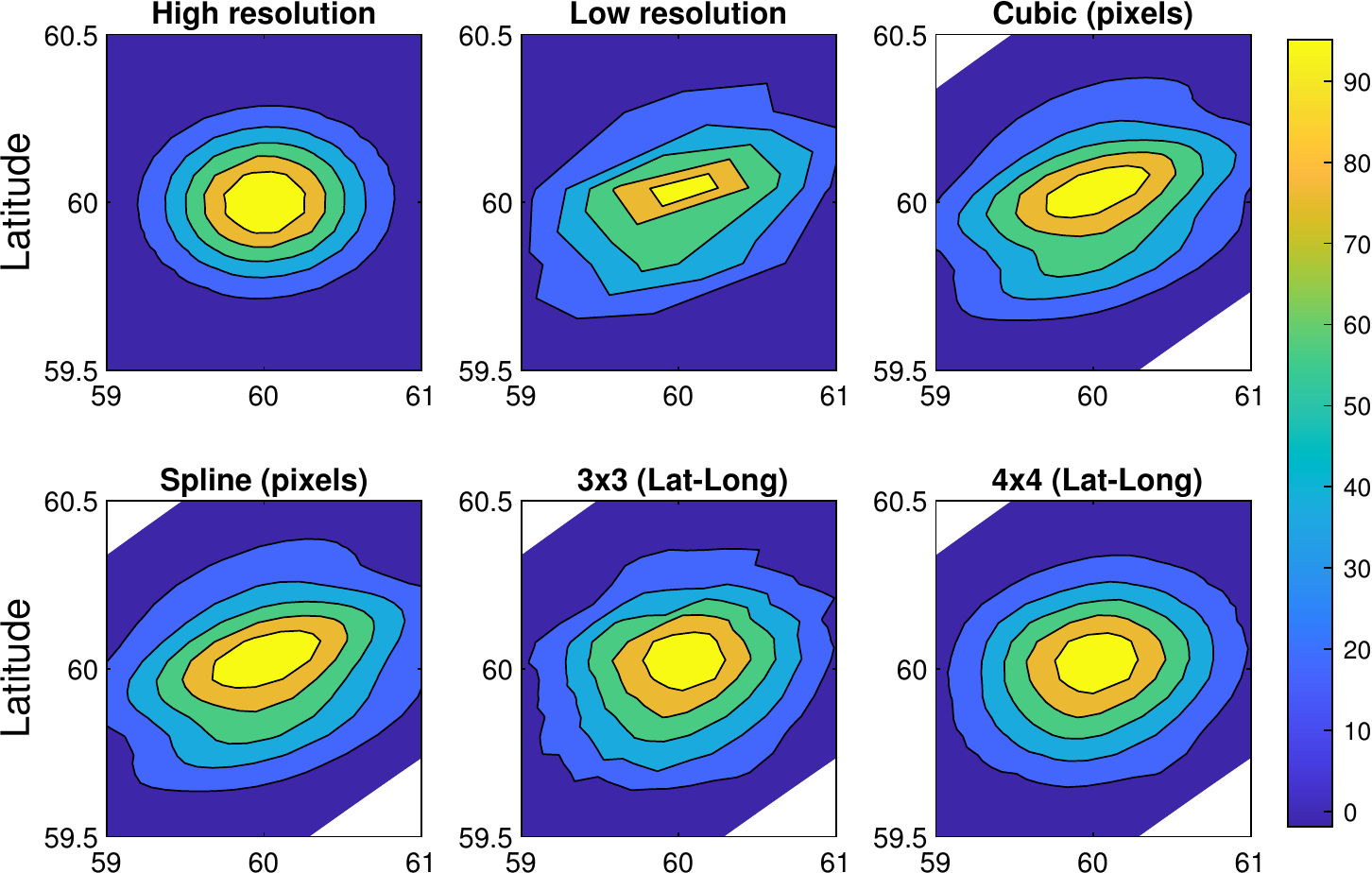}\\
    \vspace{7 mm}
    \includegraphics[width = 12 cm]{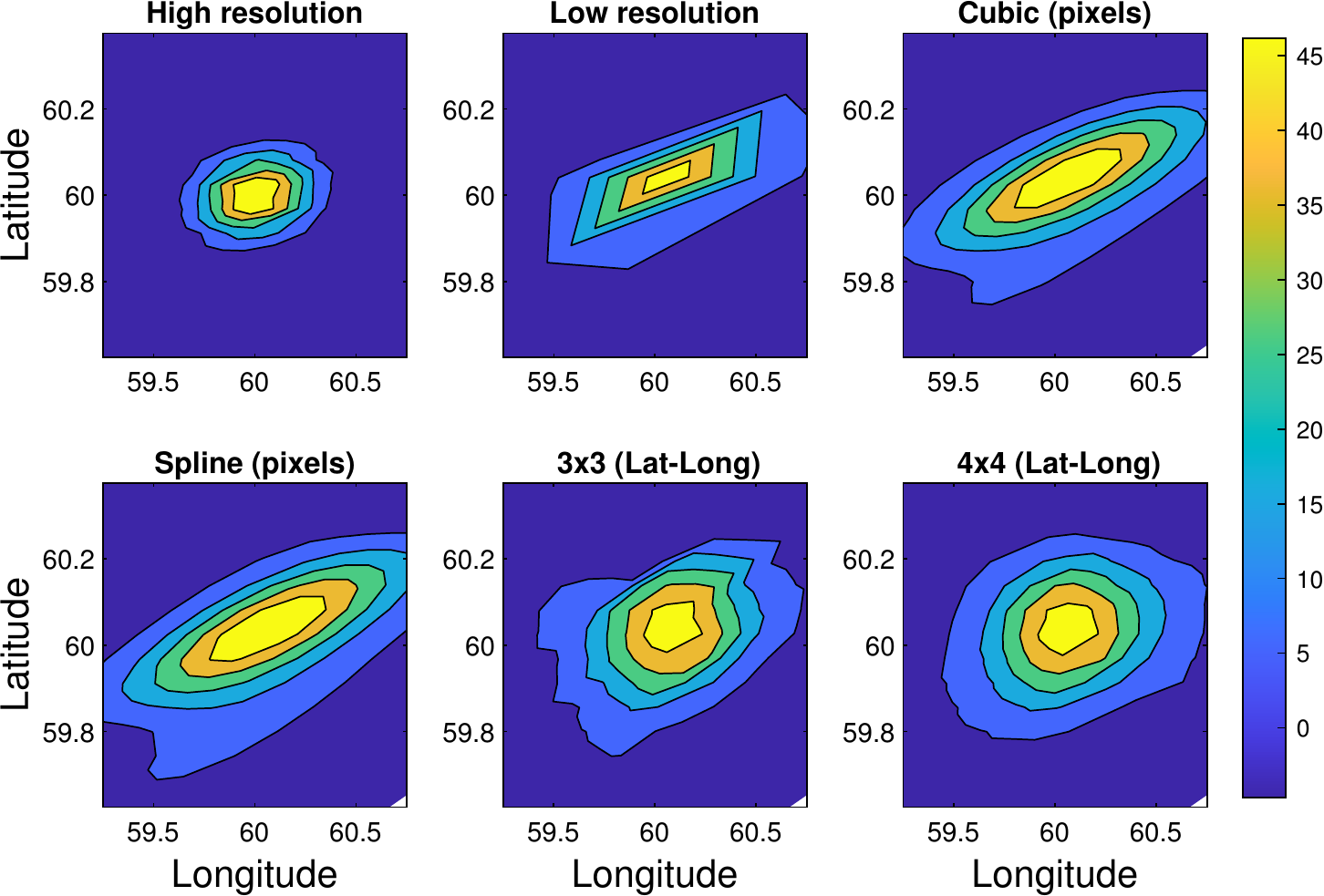}
    \caption{The performance of various interpolation algorithms used in projection routines. The first two panels in the top row show the heliographic projection of a well resolved 2D Gaussian magnetic feature analytically placed on a synthetic magnetogram and sampled at different resolutions. The third panel in the first row and all panels in the second row show the same magnetic feature when different interpolation algorithms are used to reconstruct the higher resolution projection of the feature by oversampling its low resolution image. In the third and fourth rows, corresponding plots for a feature which is barely resolved in the high-resolution image and unresolved in the low resolution image are shown.}
    \label{fig:latlong_interpolation}
\end{figure}

Most correlation tracking studies on solar data project the solar images onto a two dimensional plane before the movements of features are tracked. We choose the equirectangular Heliographic projection (uniformly spaced in latitude and longitude) for our task following the prescription of \citet{2010Sci...327.1350H}. It is extremely important to minimize uncertainty in projected locations of features and distortion in their shape to maximize the accuracy of correlation tracking. For this purpose, we have comprehensively tested our Heliographic projection algorithm using the AMC and the CCD emulator.

In Heliographic projection algorithms, it is common to interpolate solar images using a bicubic or bicubic spline interpolant (for example in \citet{2018A&A...619A..99L}) which operates on the uniformly spaced image pixel grid. However, these algorithms incorrectly assume that the physical information in each pixel is equally spaced in physical distance as well. To evaluate the accuracy of these interpolation techniques, we performed the following experiments. In two separate experiments two magnetic features, one well resolved (FWHM = 5.33 Mm) and one barely resolved (FWHM = 1.78 Mm), were placed at $60^{\circ}$ latitude and $60^{\circ}$ longitude using the AMC and each of them were then sampled by the CCD emulator at two different resolutions: $4096\times 4096$(high) and $1024\times1024$(low). Both bicubic and bicubic spline interpolation were then used to oversample the low resolution magnetograms in order to reconstruct their high resolution counterparts. We found that interpolation over the image pixel grid distorts the oversampled magnetograms as shown in Fig.\ \ref{fig:latlong_interpolation} with the smaller feature suffering more distortion. Correlation between the oversampled and the high-resolution magnetogram listed in Table \ref{table:interpolation_correlation} was used as a measure of accuracy.

To minimize distortion in the projection algorithm, we adopted the biharmonic spline interpolation algorithm of \citet{1987GeoRL..14..139S} and employed it on the non-uniformly spaced ($\lambda$, $\phi\cos(\lambda)$) coordinates of the centers of image pixels. This interpolation algorithm relies on data values in neighboring pixels and their physical distance from the target location. Two variants of this algorithm were used, one which utilizes data from a $3\times3$ pixel area and another which uses data from a $4\times4$ pixel area surrounding a target location that is in between the second and the third pixel in both dimensions. All interpolation methods magnified the barely resolved feature (Fig.\ \ref{fig:latlong_interpolation}) more than the well resolved feature.  Bicubic and bicubic spline interpolation completely changed alignment of the barely resolved feature while they only distorted the shape of the well resolved feature.  Biharmonic spline interpolation utilizing data from $4\times4$ neighboring pixels outperformed all other methods as is suggested by the higher correlation coefficient between the oversampled and the high-resolution images listed in Table \ref{table:interpolation_correlation} and minimal visual distortion in Fig.\ \ref{fig:latlong_interpolation}. This experiment was repeated at various locations on the solar disk with similar results.

In our correlation tracking algorithm, each magnetogram is projected onto a uniformly spaced latitude-longitude heliographic grid ($-90^{\circ}$ to $+90^{\circ}$ in both directions) with number of grid points equivalent to the dimension of the magnetogram itself. So MDI magnetograms are projected onto a $1024\times1024$ and HMI magnetograms on a $4096\times4096$ heliographic grid.

\begin{deluxetable}{c c c}
    \tablecaption{ Correlation coefficient between the oversampled and high resolution image when different interpolation algorithms are used to oversample a low resolution image into a high resolution image of a feature on a synthetic magnetogram. \label{table:interpolation_correlation}}
    \tablehead{\colhead {Interpolation method} & \colhead {Correlation (well-resolved feature)} & \colhead { Correlation (barely resolved feature) }}
    \startdata
Bicubic (pixels) & 0.9604 & 0.7881\\
Bicubic spline (pixels) & 0.9664 & 0.7802\\
Biharmonic spline 3 $\times$ 3 (lat-long) & 0.9763 & 0.8380\\
Biharmonic spline 4 $\times$ 4 (lat-long) & 0.9833 & 0.8463 \\
    \enddata
\end{deluxetable}

\subsection{Choosing block sizes and changing search area with latitude}\label{sec:search_area}

\begin{figure}
\begin{flushleft}
  \centering
  \includegraphics[width=12 cm]{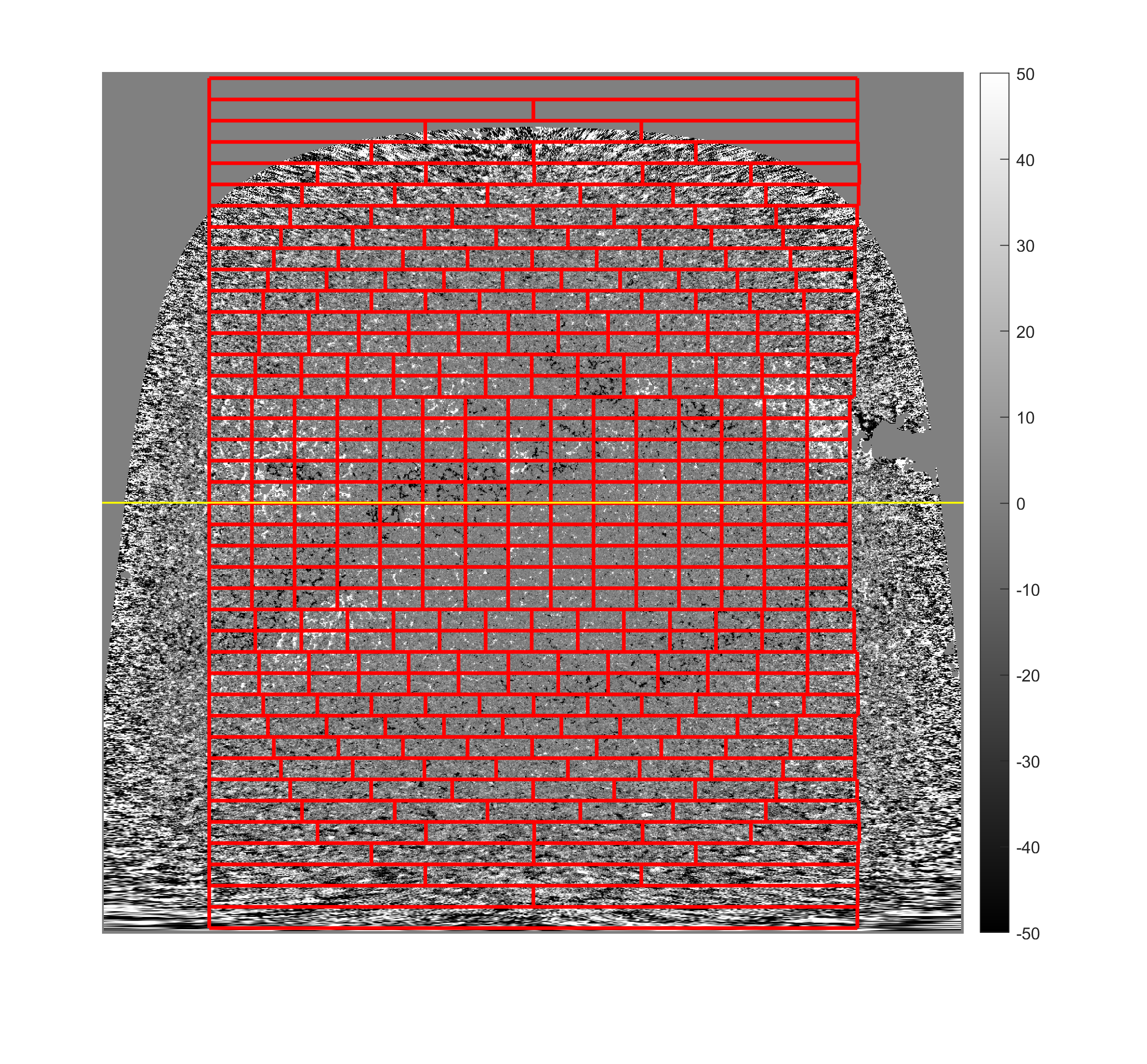}\\
  \caption{Blocks selected for tracking from an HMI magnetogram (in $Gauss$) projected onto a uniform grid in latitude-longitude are marked in red. The horizontal extent of the projected magnetogram stretches from $-90^{\circ}$ on the left to $+90^{\circ}$ longitude on the right and the vertical extent from the south pole to north pole. The yellow line indicates the solar equator.}\label{blocks}
\end{flushleft}
\end{figure}

\begin{deluxetable}{c c c}
    \tablecaption{Search area at solar disk center as a function of time-lag is shown here chosen so that velocities upto $\pm 400$ $m/s$ can be detected. The time unit for MDI is $96$ minutes and for HMI is one hour. The longitudinal search area at the equator is equal to the values in this table and varies with a factor of the secant of the latitude. \label{searcharea_table}}
    \tablehead{\colhead {No. of time units} & \colhead {HMI search area} & \colhead{MDI search area} }
    \startdata
1  & $\pm 0.132^{\circ}$  & $\pm0.352^{\circ}$ \\
2  & $\pm 0.308^{\circ}$  & $\pm0.527^{\circ}$ \\
4  & $\pm 0.659^{\circ}$  & $\pm1.055^{\circ}$ \\
8  & $\pm 1.362^{\circ}$ & $\pm2.285^{\circ}$  \\
    \enddata
\end{deluxetable}

\cite{2010Sci...327.1350H} used thin longitudinal strips to measure meridional flow in line-of-sight magnetograms. This approach, however, does not provide us any information about longitudinal variation in the measurements. We have chosen to split the projected magnetogram into rectangular blocks in latitude-longitude such that each block is large enough to contain at least one supergranule ($\approx 30 Mm$). Each of these blocks has a latitudinal extent of $4.44^{\circ}$ so that there are 40 blocks covering all latitudes. To avoid tracking a block all the way to the limb of the Sun, all blocks were chosen within $60^{\circ}$ longitude from the central meridian and their longitudinal extent was $8.93^{\circ}$ at the equator. To keep the physical area inside each block constant, their longitudinal extent was varied as a function of latitude as shown in Fig.\ \ref{blocks}. Tracking such blocks provides us less latitudinal resolution than thin longitudinal strips, but we gain longitudinal information in flow patterns. 

Each of the blocks from one magnetogram ($mag1$) was cross-correlated within a search area around their expected location (estimated by the average differential rotation) in another magnetogram ($mag2$) a certain time apart. This process was done both forward (select a block in $mag1$ and search around its expected location in $mag2$) and backwards (select a block in $mag2$ and search around its expected location in $mag1$) in time to double the number of measurements and minimize any systematic errors. This search area was varied with time-lag between successive magnetograms and by a factor of secant of the heliocentric latitude of block center (see Table \ref{searcharea_table}) in order to keep the search area constant in physical units. The search area was chosen so that it will encompass the distribution of velocities beyond three standard deviations. In post-processing, we drop velocity measurements beyond two standard deviations for each block while calculating the average velocity (averaged over all magnetogram pairs in each Carrington rotation). Averaging without excluding these measurements results in uneven coverage of the tails on either side of the distribution and leads to underestimation of the velocities by biasing the measurements towards the center of the search area.

To track the network magnetic elements, any 9$\times$9 pixel area in projected HMI or 3$\times$3 pixel area in projected MDI magnetograms with an average magnetic field in excess of 150 $Gauss$ was masked.

\subsection{Peak-locking}\label{sec:peak_locking}

\begin{figure}
\begin{flushleft}
  \centering
  \includegraphics[width=15 cm]{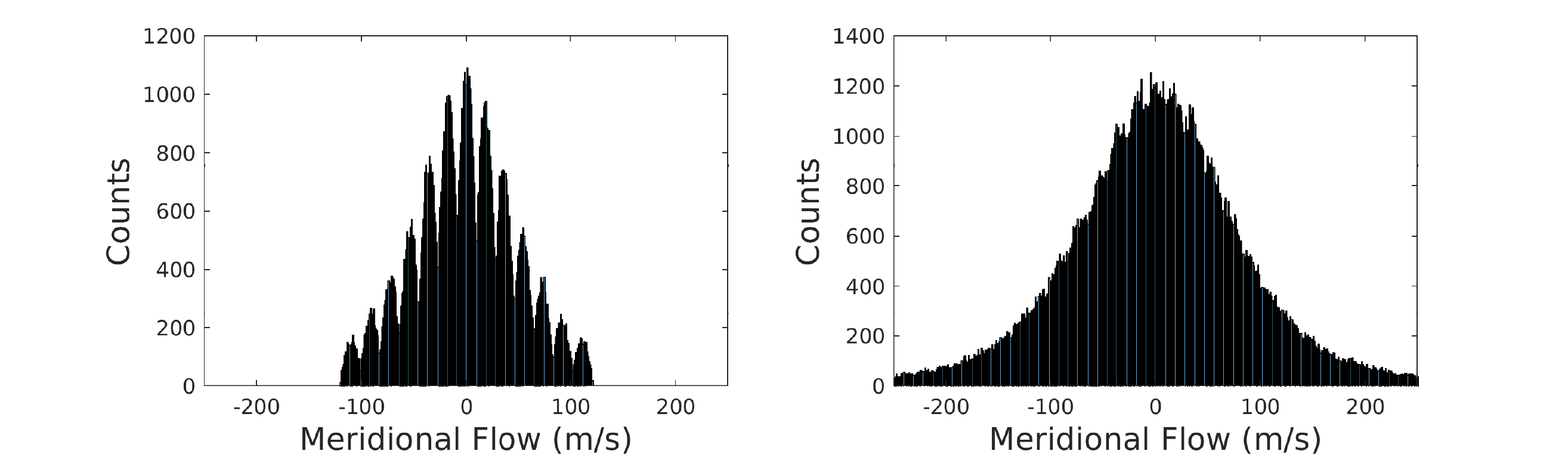}\\
  \caption{\textit{Left:} A histogram of all measurements of meridional flow velocity from Carrington rotation 2098 at 8 hr time-lag shows peak-locking at $18$ $m/s$ intervals. Moreover, the measurements in the original algorithm were clipped due to a small search area. \textit{Right:} Corresponding histogram of meridional flow measurements from our improved algorithm mitigates peak-locking and captures the full range of the velocities making both the individual and average measurements more accurate. }\label{fig_peak_locking}
  \end{flushleft}
\end{figure}

\cite{2010Sci...327.1350H} found the location of the maximum correlation by finding the parabola through the peak of the correlation coefficient to measure the displacement of each block within any time-lag. Such a parabolic fit has an accuracy of $~0.1$ pixel shift. A histogram of all measurements of meridional flow velocities at a time-lag of eight hours obtained using this method is shown in Fig.\ \ref{fig_peak_locking}. We found that a simple parabolic (or 2D Gaussian) fit to the spatial distribution of correlation coefficient results in a phenomenon known as peak-locking in the field of particle image velocimetry \citep{1997MeScT...8.1427H,1997MeScT...8.1379W,2005MeScT..16.1605C}. We get more measurements at integer pixel shifts (equivalent to every $18$ $m/s$ at HMI resolution for 8 hr time-lag) than at fractional pixel shifts. Half-pixel shifts tend to be far less likely to be found. This phenomenon appears to be known in the solar-physics community \citep{2015A&A...581A..67L} but has not been specifically addressed by name. While \citet{2015A&A...581A..67L} and B.T. Welsch (private communication) address this by an iterative procedure which translates the cross-correlation, we resolve this problem by taking the result from the parabolic fit as an estimate of the displacement, and then interpolating the reference block to shift it by this fractional pixel estimate and iterating the procedure until we find a near zero shift. The total displacement is then a sum of shifts through all iterations. Implementing such an iterative procedure improved our histogram of meridional flow measurements (see Fig.\ \ref{fig_peak_locking}) and increased the accuracy of tracking five times to $0.02$ pixel shift.

\subsection{Discovery and Correction of the Center-to-limb effect}\label{sec:center_to_limb_correction}

\begin{figure}
\textbf{Tracking On Signed Magnetograms}
\vspace{4 mm}
{    \centering
    \includegraphics[width = 18 cm]{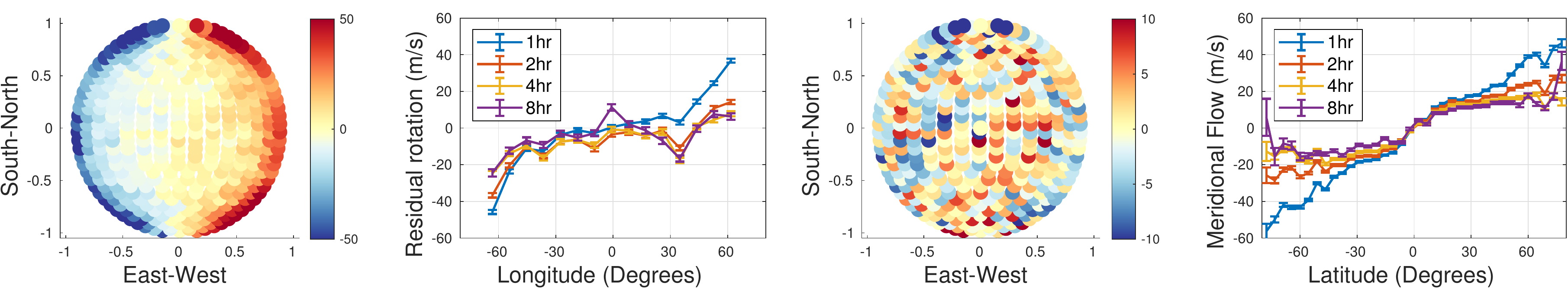}
}
\textbf{Tracking On Unsigned Magnetograms}
\vspace{4 mm}
{ \centering
    \includegraphics[width = 18 cm]{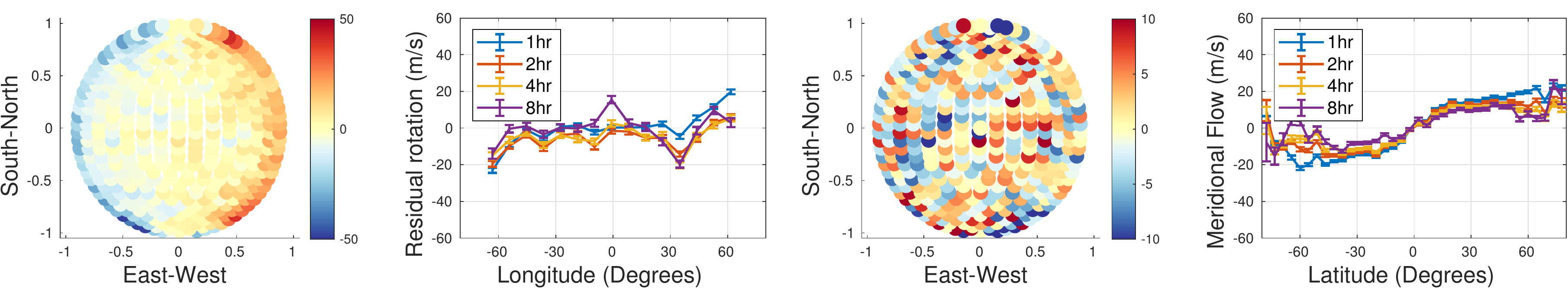}
}
\textbf{Tracking On Synthetic Magnetograms}
{ \centering
    \includegraphics[width = 18 cm]{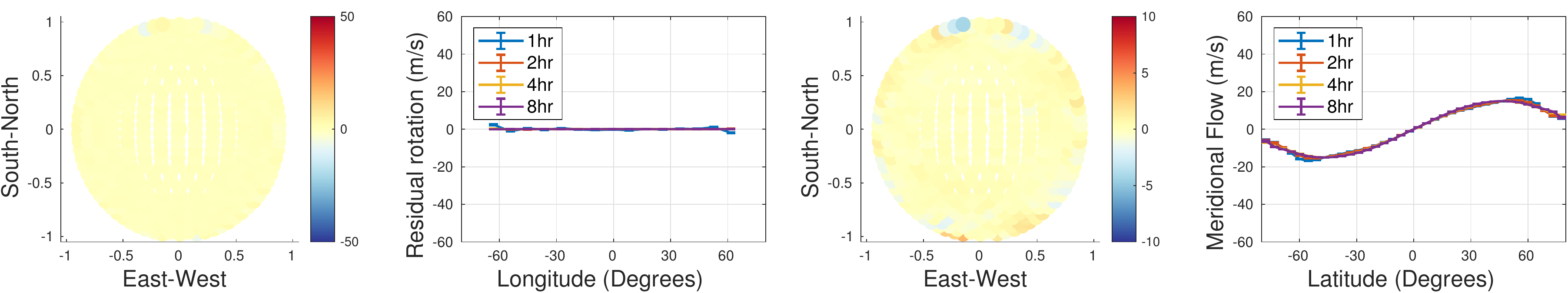}\\
    \caption{A comparison of tracking results from signed magnetograms (top row) and unsigned magnetograms (middle row) for Carrington rotation 2098. The bottom row shows tracking results from synthetic data created with a noise level of 10 Gauss (FWHM). The first column in each row shows the residual rotation rate measured at one hour time-lag obtained after removing the average rotation rate measured at each latitude, while the second column shows the residual rotation rate measured at the equator at several time-lags. The third column in each row shows the residual meridional flow speed measured at one hour time-lag obtained after removing the average meridional flow speed measured at each latitude, while the fourth column shows the latitudinal variation of meridional flow speed at several time-lags. }
    \label{fig:lev1_systematics}
}
\end{figure}

Flow measurements from correlation tracking performed on signed magnetic field at several time-lags averaged over Carrington rotation 2098 are shown in the top row of Fig.\ \ref{fig:lev1_systematics}. These measurements indicate that the east limb of the Sun is rotating $\sim 100$ m/s slower than the west limb of the Sun at one hour time-lag. This difference in rotation rate between the east and west limb decreases as a function of time-lag. The meridional flow measurements after removal of the longitudinally averaged flow speed at each latitude do not show any limb to limb variation in the longitudinal direction. The meridional flow profiles themselves, however, vary significantly as a function of time-lag. This pattern of variation with time-lag was also seen when the algorithm of \cite{2010Sci...327.1350H} was used for tracking. This combination of variation in meridional flow and the east-west asymmetry in rotation rate with time-lag is reminiscent of the center-to-limb variation reported by \cite{2012ApJ...749L...5Z}. It appears like a nearly instantaneous displacement of all features away from disk center which contributes less to velocity measurements at longer time-lags.

A very similar variation in meridional flow (measured from one month of data) was reported by \citet{2017ApJ...836...10L} by tracking individual magnetic features with varying lifetimes: 2 hr, 10 hr and 20 hr. \citet{2017ApJ...836...10L} found faster flow speeds for the shorter lived features. This variation in meridional flow speed as a function of feature lifetimes was attributed to short-lived flows and apparent motion due to feature-feature interactions. However, the longitudinal variation of the flow velocities was not analyzed in this study. The fact that we see unrealistic longitudinal variation in rotation rate along with meridional flow speed variation rules out short-lived flows as the culprit and indicates that the source of this discrepancy must be a systematic error because the Sun cannot know which meridian is centered in HMI magnetograms and cannot adjust its rotation rate accordingly.

Interestingly, this systematic center-to-limb effect reduces significantly (by $\sim 50\%$) when unsigned magnetograms are used for tracking as is shown in the second row of Fig.\ \ref{fig:lev1_systematics}. The third row of Fig.\ \ref{fig:lev1_systematics} indicates the absence of these patterns in flow speeds measured from synthetic magnetograms suggesting that these effects do not originate from our projection or tracking algorithm. In order to understand why tracking the unsigned magnetic field improves the measurements, we must take a closer look at the dynamics of the solar photosphere.

Small scale magnetic features on the photosphere of the Sun tend to accumulate near converging flows in granules and supergranules and tend to be absent at cell centers where the plasma flow is diverging \citep{1964ApJ...140.1120S}. When we use the signed value of magnetic field for correlation tracking, we track the positive and negative magnetic features themselves without giving any consideration to places where they are not found. When we use the absolute value of the magnetic field we also track cell centers as negative features when the average is removed. The reduction in the apparent shift away from disk center suggests that this shift is directly associated with the magnetic features themselves.

Even though correlation tracking on unsigned magnetograms shows a reduced systematic shift, some residual variation of meridional flow persists and the rotation rate still shows some asymmetry in the east-west direction. So, we develop a robust methodology for removing systematic errors from flow measurements which would be generally applicable regardless of whether the magnetograms are signed or unsigned and even independent of the features (magnetic, doppler, photometric, etc.) being used for tracking. 

 We begin by making certain simplifying assumptions about the composition of measured displacements of blocks. The simplest way would be to assume that there exists a fixed systematic shift, $\Delta$, away from disk center at all time-lags. With no intrinsic variation in baseline (true) flow speed ($v$) with time-lag, this constant shift affects the shorter time-lag measurements more than the longer time-lag measurements. This, however, proved to be too simple. When we decomposed the measured displacements for each block at several time-lags into a displacement due to a baseline velocity ($v\Delta t$) and a constant shift ($\Delta$) we found different $\Delta$ for different time-lags which violated our initial assumption. We then added a possible intrinsic variation in flow speed with time-lag by adding another term, $\delta$, representing a linear variation in flow speed with time-lag. We thus included:
\begin{itemize}
    \item $\Delta$ component: A constant shift independent of time-lag defined as in Eq. \ref{profile_decomposition}.
    \item $\delta$ component: A quantity with the dimensions of acceleration which accounts for changes in flow speed as a linear function of time-lag.
\end{itemize}

Using the definitions above, we can write the displacement, $D_{n}$ of a particular block, detected by correlation tracking at time-lag n$\Delta t$, where $\Delta t$ is 1-hour for HMI and 96-minutes for MDI, as

\begin{equation}\label{profile_decomposition}
    D_{n} = (v + \delta\ n \Delta t)n\Delta t + \Delta
\end{equation}

\noindent so that $\delta$ and $\Delta$ can be determined via the expressions

\begin{equation}
\delta_{1-2-4} = (D_{4} -3D_{2}+2D_{1})/(6\Delta t^{2})    
\end{equation}
\begin{equation}
\centering
\delta_{2-4-8} = (D_{8} -3D_{4}+2D_{2})/(24\Delta t^{2})    
\end{equation}

\begin{equation}
    \Delta_{1-2} = 2D_{1}-D_{2} + 2\delta\ \Delta t^{2}
\end{equation}
\begin{equation}
  \Delta_{2-4} = 2D_{2}-D_{4} + 8\delta\ \Delta t^{2}  
\end{equation}
\begin{equation}
\centering
  \Delta_{4-8} = 2D_{4}-D_{8} + 32\delta\ \Delta t^{2}.  
\end{equation}

\noindent Finally, we can derive ``baseline" velocity profiles (a velocity independent of the systematic shift and time-lag variations)
\begin{equation}
v_{1-2} = (D_{2} -D_{1})/(1\Delta t) - 3\delta\ \Delta t  \\
\end{equation}
\begin{equation}
v_{2-4} = (D_{4} -D_{2})/(2\Delta t) - 6\delta\ \Delta t  \\
\end{equation}
\begin{equation}
v_{4-8} = (D_{8} -D_{4})/(4\Delta t) - 12\delta\ \Delta t.  \\
\end{equation}

Here the subscripts, e.g. 1-2-4, indicate which time-lag displacements were used to derive the quantity, e.g. 1$\Delta$t, 2$\Delta$t, 4$\Delta$t.

\begin{figure}
    \centering
    \includegraphics[width = 18 cm]{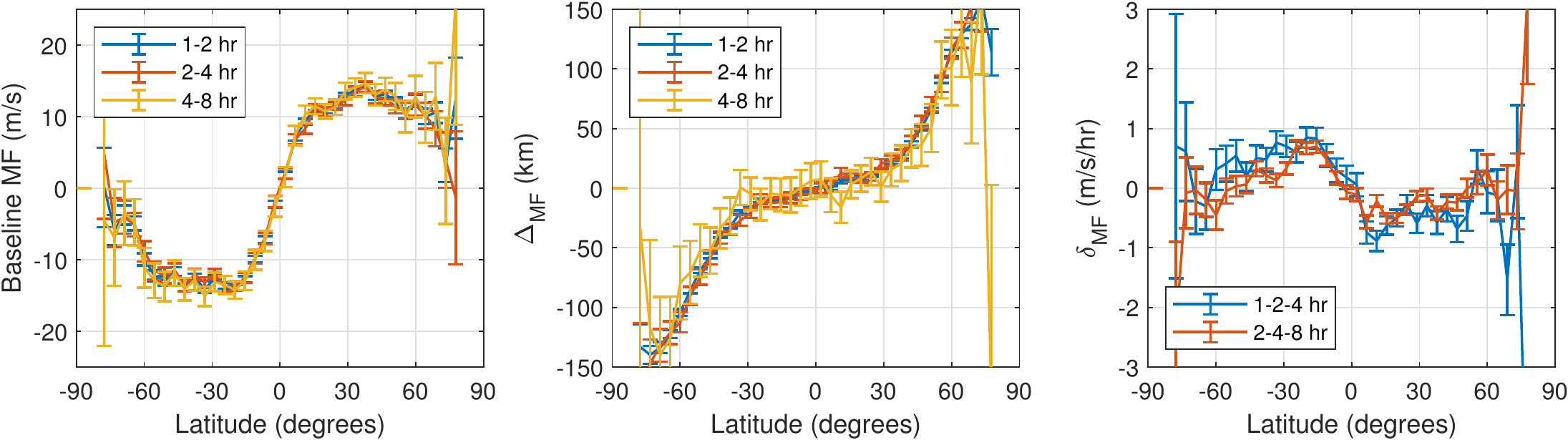}
    \caption{ From left to right: $v$, the baseline meridional flow velocity, constant shift $\Delta_{MF}$ and the time-lag dependency parameter $\delta_{MF}$ derived from combinations of measurements at different time-lags as in Eqs. 4 to 11 and averaged over one year from May 2010 to April 2011 for HMI. Their agreement supports our assumption that $\Delta$, $\delta$ and the baseline flow are independent of time-lag.}
    \label{fig:assumption_proof}
\end{figure}

\begin{figure}
    \centering
    \includegraphics[width = 18 cm]{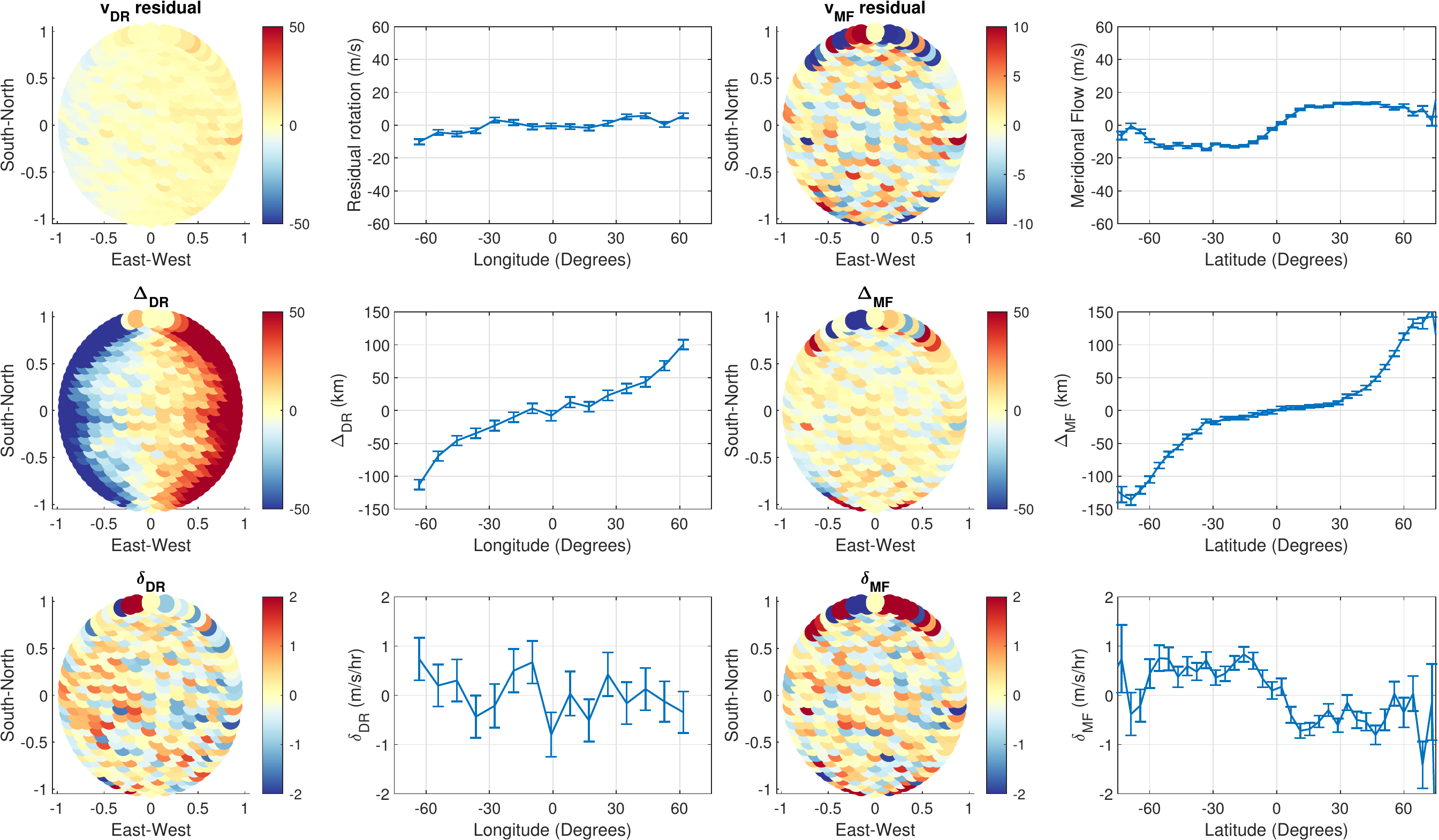}
    \caption{Baseline flow velocity $v$, $\Delta$, and $\delta$ averaged over one year (from May 2010 to April 2011) and over different combinations of time-lags. The top row shows the baseline flow velocity (in $m/s$), middle row shows $\Delta$ (in $km$) and $\delta$ (in $m/s/hr$) is shown in the bottom row. The first and third columns show the residuals after removing the average quantity at each latitude. The second column shows the longitudinal dependence of the quantity at the solar equator and the fourth column shows the latitudinal dependence (averaged over all longitudes) of the quantity.}
    \label{fig:average_systematics}
\end{figure}

The baseline velocity profiles, $v$, time-lag dependency parameter, $\delta$, and constant shift, $\Delta$, for the meridional flow and rotation rate are derived from measurements of displacements at several time-lags and averaged over each Carrington rotation. One year temporal averages of longitudinally averaged $v,\delta$ \& $\Delta$ shown in Fig.\ \ref{fig:assumption_proof} do not violate our initial assumption that these parameters are independent of the combination of time-lags used to derive them up to time-lags of 8 hr for meridional flow and 4 hr for rotation rate. The simplified assumption of $\delta$ component representing a linear variation in baseline velocity with time-lag may not hold beyond time-lags of 4 hr for rotation rate. This is thought to be due to the fact that at longer time-lags, our measurements would be biased towards tracking larger, stronger and longer-lived magnetic features that may be rooted deeper inside the Sun and hence move with the rotation rate of deeper layers which does not strictly vary linearly with depth \citep{1998MNRAS.298..543A,1998ApJ...505..390S,2009LRSP....6....1H}. 

The spatial distribution of one year averages of the derived baseline flow profiles ($v$), $\delta$ and $\Delta$ averaged over different combinations of time-lags is shown in Fig.\ \ref{fig:average_systematics}. Amazingly, the systematic center-to-limb error is automatically captured by $\Delta$ in both meridional flow and rotation rate while the baseline flow profiles and the time-lag dependency parameter do not show a significant, unrealistic East-West dependence.

\section{Results}

We now discuss the measurements obtained after making aforementioned improvements to our LCT algorithm and speculate on the physical meaning of baseline flow profiles ($v$) and that of $\Delta$ and $\delta$. The constant shift $\Delta$ shows an unrealistic East-West variation which is a clear indication that this component is a systematic error that varies as a function of position on the solar disk just like the center-to-limb effect. The time histories of the derived components ($v$, $\Delta$ \& $\delta$) are shown as functions of latitude in Fig.\ 8. Since the latitude at the center of the solar disk varies annually, we see annual variations in $\Delta$ in Fig.\ \ref{fig:composite_mdi_hmi}. If this $\Delta$ component were not removed, these annual variations would contaminate the flow velocity measurements. The time-lag dependency parameter $\delta$ however does not show any unrealistic dependence on longitude in Fig.\ \ref{fig:average_systematics} and thus is probably not a systematic error but rather an indication of variation in meridional flow and rotation rate with feature lifetime. The interpretation and properties of $\Delta$ and $\delta$ are further discussed in section \ref{sec:error_properties}.

\subsection{Meridional Flow}

The baseline meridional flow velocity averaged during the MDI and HMI overlap period from May 20, 2010 to Jan 11, 2011 (shown in Fig.\ \ref{fig:calibration_mdi_hmi}) was nearly identical (within error bars) for latitudes $<45^{\circ}$ and no cross-calibration was required to create the top left panel of Fig.\ \ref{fig:composite_mdi_hmi}. It is expected that the meridional flow measurements at higher latitudes are more trustworthy from HMI due to its higher resolution and lower noise level magnetograms. The footprint of solar active regions is clearly visible in the meridional flow panel of Fig.\ \ref{fig:composite_mdi_hmi} with meridional flow being slower (by around 5 $m/s$) at mid-latitudes during the sunspot maximum. This variation is represented in the fitting coefficient of the associated Legendre polynomial $P_2^1(\lambda) = 2\sin\lambda \cos\lambda$ to the meridional flow profile shown in the left panel of the last row of Fig.\ \ref{fig:composite_mdi_hmi} along with error bars that span two standard deviations. Note that the temporal trend in this fitting coefficient for the baseline velocity $v$ is remarkably similar to the fitting coefficient for 8-hour time-lags reported by \citet{2010Sci...327.1350H} but the amplitude in our measurements is systematically higher. This is due to the small effect of $\Delta$ and $\delta$ at 8-hour time-lags which reduces the apparent meridional flow. The fit coefficient shows that the amplitude of the meridional flow was slowest near the sunspot maximum of solar cycle 23. The top left panel of Fig.\ \ref{fig:composite_mdi_hmi} captures the spatiotemporal variations in meridional flow in greater detail.

\subsection{Torsional Oscillation}

\begin{figure}
    \centering
    \includegraphics[width = 10 cm]{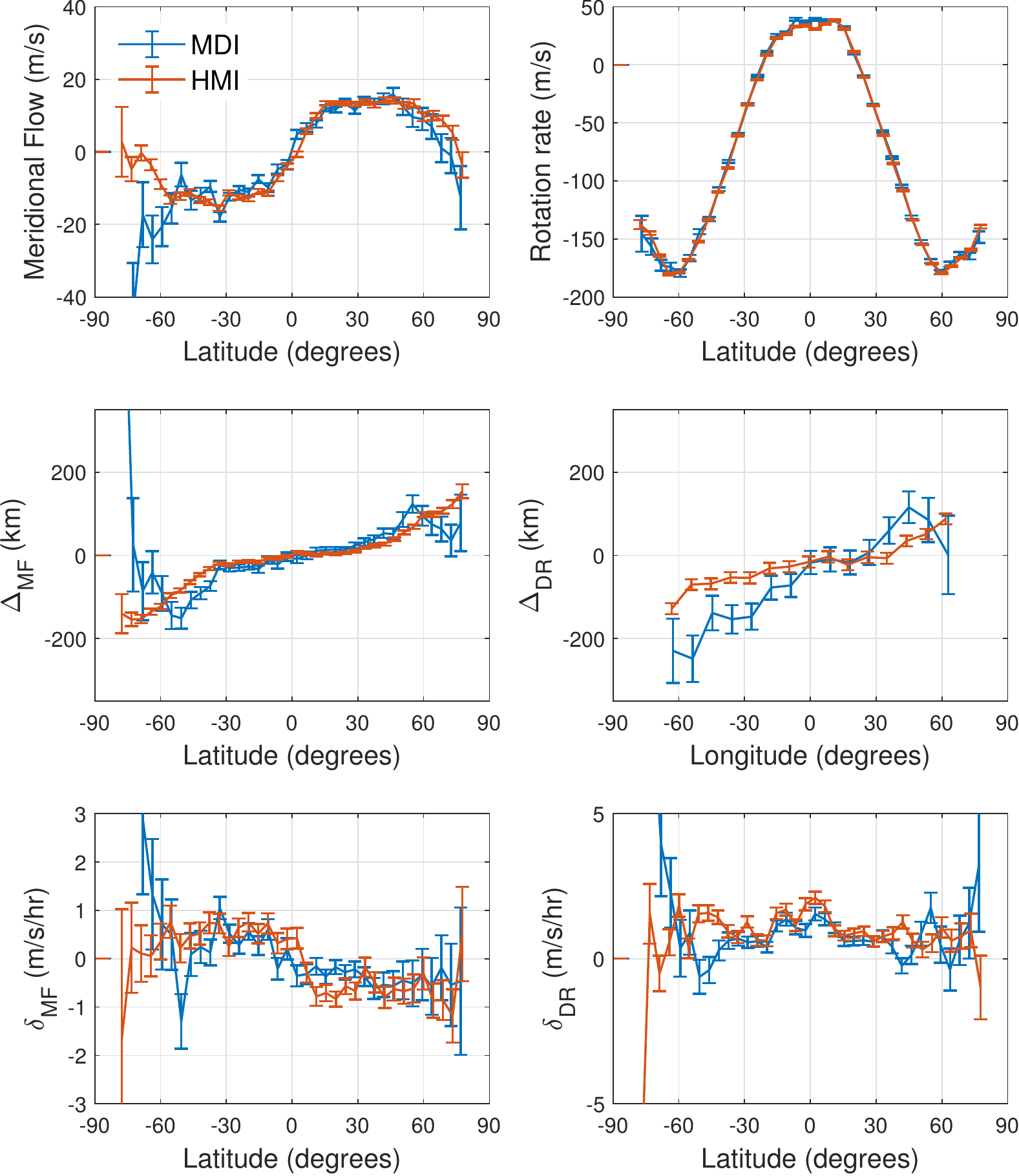}
    \caption{ Measured quantities temporally averaged during the overlap period between MDI (blue) and HMI (red). \textit{Top row:} baseline meridional flow and rotation rate (in Carrington frame). \textit{Middle row:} center-to-limb error in MF and DR. \textit{Bottom row:} time-lag dependency parameter $\delta$ in MF and DR. The horizontal axis for $\Delta_{DR}$ is longitude near the equator. }
    \label{fig:calibration_mdi_hmi}
\end{figure}

The baseline rotation rate measured in MDI magnetograms during the overlap period (shown in Fig.\ \ref{fig:calibration_mdi_hmi}) also was nearly identical to the one measured from HMI magnetograms except within $10^{\circ}$ of the equator. Above $50^{\circ}$ latitude, rotation rates obtained from HMI were much less noisy (by a factor of three) compared to MDI. 

\begin{figure}
    \centering
    \includegraphics[width = 18 cm]{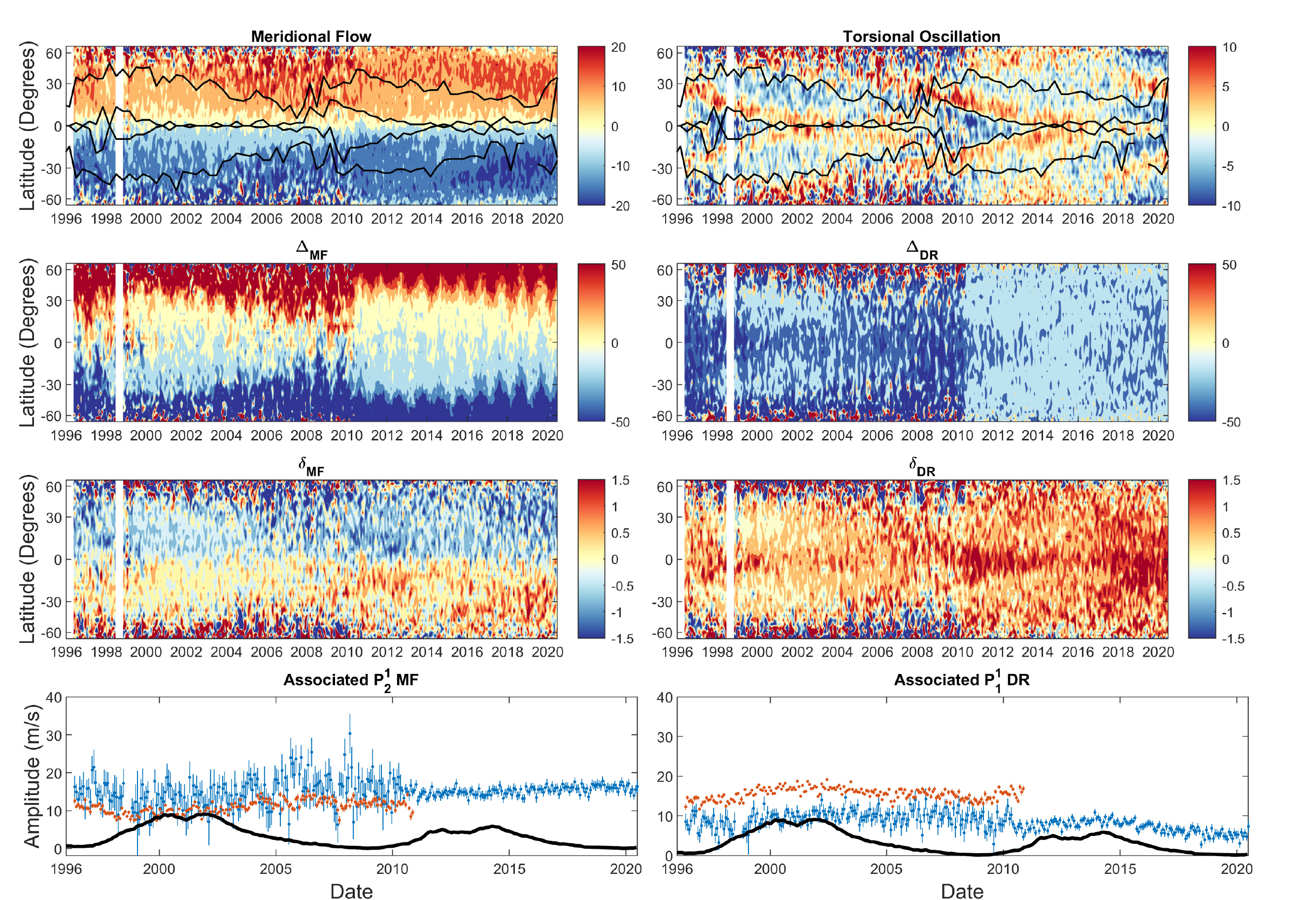}
    \caption{ Latitude-time plots of corrected flow velocities and systematic error along with temporal variation of Legendre coefficients fitted to the flow velocities. The top row shows latitude-time plot of the baseline meridional flow and torsional oscillation obtained from over 23 years of measurements from MDI and HMI data in units of $m/s$. The boundaries of active latitudes are marked in black using the Royal Greenwich Observatory sunspot records. The corresponding plots for $\Delta$ (in $km$) and $\delta$ (in $m/s/hr$) are shown in the second and the third row respectively. The bottom row shows the scaled sunspot number (in black) and our fitting coefficients of associated Legendre polynomial of order 1, degree 2 to the meridional flow profile and of order 1, degree 1 to the torsional oscillation (translated by 55 $m/s$) with $2\sigma$ errors in blue whereas the fitting coefficients from \citet{2010Sci...327.1350H} are shown with red dots and standard error. }
    \label{fig:composite_mdi_hmi}
\end{figure}

We subtracted the average rotation rate during the HMI era from HMI measurements and the average rotation rate during MDI era from MDI measurements to derive the torsional oscillation profile (deviations from the average rotation rate) in Fig. \ref{fig:composite_mdi_hmi}. This measured torsional oscillation pattern is remarkably similar to the one obtained from helioseismic techniques \citep{2000ApJ...533L.163H,2000ApJ...541..442A,2002Sci...296..101V,2009LRSP....6....1H} and shows the equatorward propagating branches of faster and slower than average rotation rate in great detail while the signal to noise ratio is not high enough to see the poleward propagation of high-latitude branches clearly. The equatorward propagating branch with faster than average rotation rate (in red) which the active regions of solar cycle 25 (expected to start from year 2020) are expected to follow is very clearly visible before the end of solar cycle 24 centered around $40^{\circ}$ latitude from late 2017 onwards and it appears disconnected from what looks like a weak patch of faster than average rotation rate at latitudes above $50^{\circ}$ preceding it.

\subsection{Properties of constant shift error $\Delta$ and time-lag dependency $\delta$}\label{sec:error_properties}

Interestingly, most of the differences between the displacement measurements ($D_{\Delta t}$) from MDI and HMI (see Fig.\ \ref{fig:calibration_mdi_hmi}) were captured by the constant shift ($\Delta$) i.e. the systematic center-to-limb error which improved the consistency of measurements across instruments. These differences may be due to dependence on certain properties of the different spectral lines which are observed to create the magnetograms e.g.: their height of formation in the solar atmosphere \citep{2016A&A...590A.130L}. The fact that we are using different sets of timelags (multiples of 96 minutes for MDI, multiples of 1 hr for HMI) may also be responsible for some part of the difference.

Most of the annual variations in uncorrected displacements were also captured by $\Delta$, leaving behind purified baseline meridional flow and torsional oscillation profiles shown in Fig.\ \ref{fig:composite_mdi_hmi}. The center-to-limb effect ($\Delta$) is defined in Eq.\ \ref{profile_decomposition} and can be re-written as
\begin{equation}
    lim_{\Delta t \to 0} D_{\Delta t} = \Delta
\end{equation}
which means it is a constant shift away from disk center that would be there even if we correlate magnetograms with near zero time-lag. When this constant shift is divided by time-lag, it dominates the measured velocities at small time-lags. For HMI at $60^{\circ}$ from disk center, we find a constant shift of $\approx 100$ $km$ in both the latitudinal and longitudinal displacements (see Fig.\ \ref{fig:average_systematics}). This implies an apparent velocity of $27.8$ $m/s$ at 1 hr and $3.5$ $m/s$ at 8 hr time-lags. While the contribution from the constant shift error decreases as a function of time-lag, the contribution from time-lag dependency parameter $\delta$ increases due to the $\Delta t^{2}$ factor.

The footprint of active regions is clearly visible in the long-term measurements of $\Delta_{MF}$ and $\Delta_{DR}$ in Fig.\ 
 \ref{fig:composite_mdi_hmi} suggesting that this systematic error is higher in the quiet Sun regions than in the vicinity of active regions. This explains why \cite{2017ApJ...836...10L} measured a very high meridional flow for weak features with life times up to two hours which primarily are found in quiet Sun regions. \cite{2016A&A...590A.130L} also suggested that the center-to-limb effect may depend on the magnetic field which agrees with our observation here. Contamination from this constant shift/center-to-limb error may also be the cause of deviation between the actual flow measurements and the fitted functional form of the meridional flow profile shown in Fig.\ 3 of \citet{2018ApJ...864L...5I}.

The time-lag dependency parameter, on the other hand accounts for variation in flow velocity with time-lag so that the effective flow velocity at any time-lag ($v'_{\Delta t}$) is slightly different from the baseline flow velocity as
\begin{equation}\label{eq:actual_velocity}
    v'_{\Delta t} = v + \delta \Delta t.
\end{equation}

The patterns of time-lag dependency parameter in meridional flow $\delta_{MF}$ and rotation rate $\delta_{DR}$ shown in Fig.\ \ref{fig:composite_mdi_hmi} show some abrupt change between the two instruments (see Fig.\ \ref{fig:calibration_mdi_hmi}). $\delta_{MF}$ hovers around $\sim 0.6$ $m/s/hr$ with a sign opposite to that of the mean meridional flow in each hemisphere indicating that the meridional flow slows down as a function of time-lag. Whereas $\delta_{DR}$ is in the range $0.0$ to $2.0$ $m/s/hr$ over most of the solar disk indicating that the rotation rate increases with time-lag. The one year average of $\delta_{MF}$ and $\delta_{DR}$ plotted in Fig.\ \ref{fig:average_systematics} does not show any unrealistic longitudinal variation and appears to be axisymmetric. Therefore, there is no convincing reason to suspect that $\delta_{MF}$ and $\delta_{DR}$ are some kind of systematic errors.

\section{Discussion \& Conclusion}

While it is clear that the constant shift ($\Delta$) comprises of the center-to-limb error, there isn't a convincing, straightforward physical interpretation of time-lag dependency parameter ($\delta$). We speculate that $\delta_{DR}$ accounts for a physical increase in rotation rate with increase in time-lag. At longer time-lags the measurements are biased towards tracking larger, stronger magnetic features with longer lifetimes that are anchored deeper inside the Sun and hence move with the velocities of the deeper layers as suggested by \citet{2012ApJ...760...84H}. The behavior of $\delta_{DR}$ also agrees with helioseismic inversions of rotation rate which show a near surface shear layer where the rotation rate increases as we look deeper below the photosphere \citep{1998MNRAS.298..543A,1998ApJ...505..390S,2008ApJ...681..680A,2009LRSP....6....1H}.

On the other hand, for $\delta_{MF}$, our measurements suggest that meridional flow decreases with depth below the photosphere. This was first noted by \citet{2012ApJ...760...84H} in the measured meridional motions of supergranules which indicated that the poleward meridional flow seen at the photosphere decreases with depth across the surface shear layer and becomes equatorward at the base of this layer and this was confirmed by helioseismic studies \citep{2013ApJ...774L..29Z,2015ApJ...805..133J,2017ApJ...849..144C} and \citet{2018ApJ...864L...5I} as well. However, \citet{2010ApJ...717..488B} found that the meridional flow speed actually increased with depth in the outer $2\%$ of the Sun. Even though the flow velocities derived in the above mentioned studies may be affected by the center-to-limb effect, their qualitative conclusion (rotation rate increases with field strength and depth; meridional flow decreases with field strength and depth) agrees with ours.

In the light of our speculation, the visibility of the footprint of active regions in the pattern of $\delta_{MF}$ and $\delta_{DR}$ then suggests that the difference between the velocities of the deeper layers and the photosphere is less in the presence of high magnetic field strength features. This, again, is explained by the fact that stronger magnetic features which are rooted deeper below the photosphere but appear at the photosphere to move with the velocity of the deeper layers in which they are anchored, reduce the apparent difference in the velocity between the photosphere and deeper layers in their neighborhood.

If our speculation on the meaning of $\delta_{DR}$ and $\delta_{MF}$ is correct then the implied meaning of the baseline flow velocities is that: this is the velocity with which the weakest magnetic features move and the stronger magnetic features (predominantly tracked at longer time-lags) move with $v'_{\Delta t}$ (Eq.\ \ref{eq:actual_velocity}).

Regardless of whether our speculations are correct, the improvements described in sec \ref{sec:improvements} are generally applicable to many other tracking algorithms and they have increased the accuracy of our LCT algorithm five fold so that it can detect a $0.02$ pixel shift reliably. Our proposed methodology for the removal of the center-to-limb effect is computationally more expensive than the traditional approach (removing some constant profile) as it utilizes measurements of displacements at several time-lags but it works with very few assumptions. To use it effectively, one does not have to assume that the center-to-limb effect is symmetric across the entire solar disk, that it is independent of the distribution of magnetic flux on solar disk, or that it is constant in time \citep{2012ApJ...749L...5Z,2015SoPh..290.1081K}. The variation of center-to-limb error with the phase of the solar cycle seen in Fig.\ \ref{fig:composite_mdi_hmi} clearly indicates that the center-to-limb effect cannot be assumed to be constant in time and it has to be calculated synchronously with the flow profiles. If it is assumed to be a constant, its temporal variation may contaminate measured flow profiles.

All our final baseline velocity, $\Delta$ and $\delta$ measurements as well as the Legendre fitting coefficients to baseline flow profiles shown in Fig.\ \ref{fig:composite_mdi_hmi} are available in an easy to use form at \url{https://dataverse.harvard.edu/dataverse/lct-on-solar-magnetograms} \citep{DVN/I0OWHG_2020,DVN/0IZBHY_2020}.



\acknowledgements
{The authors were supported by grant NNX14AO83G from the NASA Heliophysics Grand Challenges Program. The MDI data used are courtesy of the ESA-NASA/SOHO and the MDI science teams. The HMI data used are courtesy of the NASA/SDO and the HMI science teams. S.S.M. is grateful to NASA Ames Research Center for hosting him during the summer of 2016. The authors are also grateful to Lisa Upton for her insights and discussions that helped this paper.}

\bibliographystyle{apj}
\bibliography{references}



\end{document}